\documentstyle{utap}
\journalid{97}{20}{1997}
\articleid{1}{13}

\def\beq{\begin{equation}}
\def\eeq{\end{equation}}
\def\beqar{\begin{eqnarray}}
\def\eeqar{\end{eqnarray}}

\def\fun#1#2{\lower3.6pt\vbox{\baselineskip0pt\lineskip.9pt
  \ialign{$\mathsurround=0pt#1\hfil##\hfil$\crcr#2\crcr\sim\crcr}}}

\def\gmc{g cm$^{-3}$}
\def\kms{km s$^{-1}$}
\def\ms{$M_\odot$}
\def\ni{$^{56}$Ni}

\def\e#1{$\times$ $10^{#1}$ }
\def\ee#1{$10^{#1}$ }
\def\ltsima{$\; \buildrel < \over \sim \;$}
\def\ltsim{\lower.5ex\hbox{\ltsima}}
\def\gtsima{$\; \buildrel > \over \sim \;$}
\def\gtsim{\lower.5ex\hbox{\gtsima}}

\def\h0{$H_0$}
\def\ome{$\Omega_{\rm M}$}
\def\omela{$\Omega_\Lambda$}
\def\ll#1{$L_{\rm #1}$}
\def\mm#1{$M_{\rm #1}$}
\def\vv#1{$v_{\rm #1}$}

\title	{Type Ia Supernovae: Their Origin and \\ Possible 
Applications in Cosmology}
{Type Ia Supernovae: Their Origin and Possible 
Applications in Cosmology}

\author{Ken'ichi Nomoto, Koichi Iwamoto, Nobuhiro Kishimoto \\
Department of Astronomy, University of Tokyo, 
Bunkyo-ku, Tokyo 113, Japan}
{Nomoto et al.}

\comments{To appear in Science, 1997, May 30}

\abstract{ 
Spectroscopic and photometric evidence indicates that Type Ia
supernovae (SNe Ia) are the thermonuclear explosions of accreting
white dwarfs.  However, the progenitor binary systems and
hydrodynamical models for SNe Ia are still controversial.  The
relatively uniform light curves and spectral evolution of SNe Ia have
led to their use as a standard candle for determining cosmological
parameters, such as the Hubble constant, the density parameter, and
the cosmological constant.  Recent progress includes the calibration
of the absolute maximum brightness of SNe Ia with the Hubble Space
Telescope, the reduction of the dispersion in the Hubble diagram
through the use of the relation between the light curve shape and the
maximum brightness of SNe Ia, and the discovery of many SNe Ia with
high red shifts.
}

\keywords{
hydrodynamics
--- instabilities
--- stars: evolution
--- stars: supernovae
--- supernovae: individual (SN 1993J)
}

\begin{document}

\noindent
     {\Large S}upernovae are classified spectroscopically as Type I if
they have no hydrogen lines in their optical spectra and Type II if
they have hydrogen lines in their optical spectra.  Type I supernovae
(SNe I) are further subclassified into types Ia, Ib, and Ic on the
basis of spectra observed early in their explosion (early-time
spectra) (\cite{Filippenko-97}).  SNe Ia are characterized by the
presence of a deep Si II absorption line near wavelength 6150 \AA
(Fig. 1), and their late--time spectra are dominated by strong blends
of Fe emission lines.  SNe Ib and Ic, in contrast, do not show this Si
line.  Moderately strong He I lines, especially at 5876 \AA,
distinguish SNe Ib from SNe Ic in early-time spectra; that is, SNe~Ib
exhibit absorption lines of He I, whereas these lines are weak or
absent in SNe Ic (\cite{Filippenko-97}).  SNe II, Ib, and Ic are now
generally thought to result from the explosion of massive stars---SNe
II from single stars and SNe Ib and Ic from binary stars.  There are
spectroscopic and photometric indications that SNe Ia originate from
white dwarfs that are composed of C + O with strongly degenerate
electrons and have accreted sufficient mass from a companion to
trigger an explosion.

     However, the progenitor systems and hydrodynamical models for SNe
Ia are still controversial. Many issues need to be resolved including
(i) double degenerate(DD) versus single degenerate(SD) scenarios, that
is, whether the companion of the white dwarf is also an
electron-degenerate white dwarf or is a non-degenerate [main sequence
(MS) or evolved red giant] star; (ii) Chandrasekhar mass (Ch)
(\cite{Chandra}) versus sub-Chandrasekhar mass (sub-Ch) models; and
(iii) the explosion mechanism in the Ch models.  The answers to these
questions could lead to improved understanding of the complicated
evolution of close binaries as well as the physics of thermonuclear
explosion (\cite{Ruiz-etal-97}).  The relatively uniform light curves
and spectral evolution of SNe Ia have led to their use as a standard
candle to determine the Hubble constant ($H_0$), the density parameter
(\ome), and the cosmological constant (\omela)
(\cite{Branch-Tammann-92}).  SNe Ia are important standard candles
because they are bright enough to be observed out to a redshift ($z$)
of $\sim$ 1.  Variations of light curves and spectra among SNe Ia have
recently received attention.  SNe Ia with higher maximum brightnesses
tend to show a slower decline in their light curves
(\cite{Phillips-93}, \cite{Riess-etal-95}; Fig. 2).  This review
summarizes our current understanding of SNe Ia and focuses on these
controversial issues.

\vskip 0.5cm

\centerline{\large\sf Progenitors}

\vskip 0.5cm

\noindent
     White dwarfs composed of C+O are formed from intermediate mass
stars ($M <$ 8 \ms, where \ms~ is the mass of our sun), undergo
cooling, and eventually become dark matter, as they evolve towards
fainter luminosities.  In a close binary system the white dwarf
evolves differently because the companion star expands to transfer
matter to the white dwarf; the accreting white dwarfs are rejuvenated
and in certain cases undergo thermonuclear explosions to give rise to
SNe Ia.  Theoretically the Ch white dwarf models and the sub-Ch models
have been considered to explain the origin of SNe Ia
(\cite{Renzini-96}).  Various evolutionary scenarios have been
proposed, including (i) merging of double C+O white dwarfs with a
combined mass exceeding the Ch limit (a DD scenario)
(\cite{Iben-Tutukov-84}) and (ii) accretion of H or He by mass
transfer from a binary companion at a relatively high rate (an SD
scenario) (\cite{Renzini-96}, \cite{Nomoto-82a}).

     The DD-Ch scenario is favored from the theoretical estimate of
frequencies of the occurrence of SNe Ia.  This has stimulated a search
for DD systems, but few such systems have been discovered, and their
combined mass is smaller than Ch (\cite{Renzini-96}).  For an SD
scenario, the accreting white dwarf undergoes H burning near the
surface, which increases or decreases the white dwarf mass depending
mainly on the accretion rate, $\dot M$ (\cite{Nomoto-82a}).  Ch mass
white dwarfs can be formed with a relatively high accretion rate such
as $\dot M$ $\approx$ \ee{-8} to \ee{-6} $M_\odot$ year$^{-1}$ because
of relatively small mass ejection after H or He shell burning.
Recently, it was found that this range of $\dot M$ can be extended to
a rate faster than \ee{-6} $M_\odot$ year$^{-1}$
(\cite{Hachisu-etal-96}).  With such a high rate, the white dwarf
undergoes wind mass loss without expanding its radius, thereby
increasing its mass by steady H burning.  The SD-sub-Ch
scenario---that is, the explosion of sub-Ch white dwarfs---is a
possible outcome of accretion with 4 \e{-8} $M_\odot$ yr$^{-1}$ \gtsim
$\dot M$ \gtsim \ee{-9} $M_\odot$ year$^{-1}$.  In this case, the
ignited He shell flash is strong enough to initiate an off-center He
detonation (\cite{Nomoto-82b}, 12), which induces various types of
explosions.

     Promising candidates for the white dwarf's companion stars for
the SD scenario are (i) stars with MS masses of 1 to 1.5 \ms, which
fill the Roche lobe when they evolve through red-giants, and (ii) the
2 to 3 \ms~ stars, which undergo mass transfer near the MS.  These
cases of relatively fast accretion onto the white dwarf can correspond
to super-soft x-ray sources (\cite{vanden-etal-95}), symbiotic stars
(\cite{Renzini-96}), or both.

     To discriminate among the SD-sub-Ch, SD-Ch, and DD-Ch scenarios,
we used photometric and spectroscopic diagnostics.  If any H or He is
detected, the DD model could be ruled out.  Although high-velocity H
has not been observed in any SNe Ia, the upper limit to the H
abundance ($\sim$ \ee{-4} \ms) is still too high to rule out the SD
models.  The H-rich materials in the companion star may be engulfed in
the exploding material; thus the detection of low-velocity H could be
critical (\cite{Ruiz-etal-93}).

\vskip 0.5cm

\centerline{\large\sf Explosion Models}

\vskip 0.5cm

\noindent
     Once explosive nuclear burning is ignited, it induces
thermonuclear explosion of the white dwarf.  The outcome of this
explosion depends on how the nuclear flames and/or shock waves
propagate for the Ch and sub-Ch mass white dwarfs.  The physics
involved in these processes is rather complex, and multi-dimensional
simulations have been conducted to understand these processes
(\cite{Arnett-96}).  The results are still preliminary, and so we
focus here on those models that can account for the basic features of
SNe Ia, namely, an explosion energy of $\sim$ \ee{44} J, the synthesis
of a large amount of \ni~ (0.4 to 1 \ms), and the production of a
substantial amount of intermediate-mass elements at expansion
velocities of $\sim$ 10,000 \kms~ near the maximum brightness of the
SN explosion.

\noindent
{\em Chandrasekhar mass models}: Carbon is ignited in the central
region of the white dwarf when the central density exceeds $\sim$ \ee9
g cm$^{-3}$.  Because of strong electron degeneracy, C burning is so
explosive that it incinerates the material into Fe peak elements
(\cite{Fe-peak}).  Afterwards, the explosive nuclear flame propagates
outwards.  The flame front is subject to various types of
instabilities, including thermal instabilities, the Landau-Darrius
instability, the Rayleigh-Taylor instability, and the Kelvin-Helmholtz
instability (\cite{Niemeyer-96}).  The flame speed depends on the
development of these instabilities and the resulting turbulence,
requiring further extensive simulations on large and small scales as
well as suitable modeling of the turbulence.  Behind the flame front,
materials undergo explosive nuclear burning of Si, O, Ne, and C.  The
nucleosynthesis products depend mainly on peak temperatures, which in
turn depend on the densities encountered by the flame.  For densities
from \ee{10} to \ee6 \gmc, the products range from Fe-peak elements
(mostly \ni) to intermediate-mass elements (Ca, Ar, S, and Si),
O-Ne-Mg, and C+O.

     During the subsonic propagation of the deflagration wave, the
densities of the whole white dwarf are decreasing because of
expansion.  Therefore, the densities encountered by the flame are
determined by its speed, which is still uncertain.  Several plausible
models have been presented with one-dimensional codes
(\cite{Nomoto-82b}, 12).  In carbon deflagration models, such as model
W7 (\cite{Nomoto-etal-84}), the average flame speed is as high as
one-fifth of the sound speed.  A sequence of nucleosynthesis reactions
produces \ni, Ca-S-Si, O-Ne-Mg, and C+O behind the deflagration wave.
Because of fast propagation of the flame, the transition to the
detonation could be induced by shock compression in the outer low
density layers, which would produce some variations of \ni~ mass and
distribution, as seen in peculiar SNe Ia such as SN1991T (Fig.1).  In
contrast, delayed detonation models assume that the early propagation
of the deflagration is as slow as a few percent of the sound speed,
producing $\sim$ 0.1 \ms~ of \ni, and hence the transition from
deflagration to detonation could occur at densities of 1 \e7 to 3 \e7
\gmc.  In this case, the bulk of the white dwarf has expanded to lower
densities so that the detonation wave synthesizes Fe-peak elements and
intermediate mass elements (\cite{Khokhlov-91a}).  In the pulsating
delayed detonation model, the transition to detonation is assumed to
occur near the maximum compression (\cite{Khokhlov-91b}).  Possible
variations of the transition density could produce variations of the
\ni~ mass produced.

\noindent
{\em Sub-Chandrasekhar mass models}: A central C detonation can be
initiated by a He detonation-induced shock wave (12, \cite{Livne-90}).
Merging of double white dwarfs could ignite a central C detonation
(\cite{Shigeyama-etal-92}).  If the white dwarf mass is less than 1.07
\ms, the densities of the white dwarf matter encountered by the
detonation wave can be suitable to produce sufficient amounts of \ni~
and Si-Ca-O.  Variations of the white dwarf mass could cause the
variations of the light curves of the resultant SN Ia.

     Constraints on still uncertain parameters in these models, such
as the central ignition density, the flame speed, and the
deflagration-detonation transition density, can be provided by
comparisons of theoretical spectra and light curves with observations
and by comparisons of nucleosynthesis consequences with solar isotopic
ratios.  A combination of processes from various models may be
responsible for the variations in characteristics of the observed SNe
Ia.

\vskip 0.5cm

\centerline{\large\sf Spectra and Light Curves}

\vskip 0.5cm

\noindent
{\em Spectra}: Spectra of SNe Ia are generally homogeneous but show
some important variations.  Because SNe Ia do not have a thick H-rich
envelope, elements newly synthesized during the explosion can be
observed in the spectra.  Thus, comparison between the synthetic
spectra and observations is a powerful diagnostic of the dynamics and
nucleosynthesis suggested by the models.  Non-local thermodynamical
equilibrium(LTE) spectra have been calculated and can be used to
compare with observations.  A typical example (Fig. 3) shows agreement
between the C deflagration model (unmixed W7) and the observed optical
spectra of SNe 1992A and 1994D (\cite{Nugent-etal-97}).  The material
velocity at the stellar surface near maximum light is $\sim$ 10,000
\kms and the spectral features are identified as those of Fe, Ca, S,
Si, Mg, and O.  This implies that the abundance distribution in the
velocity space may be similar to W7.  The synthetic spectra for the
sub-Ch models seem to be less satisfactory (\cite{Nugent-etal-97},
\cite{Hoflich-Khokhlov-96}).

     Regarding the heterogeneity in the spectra, SNe 1991T (Fig. 1)
and 1991bg are the two extreme examples that have clearly revealed the
presence of spectroscopically peculiar SNe Ia (\cite{Mazzali-95}).
The pre-maximum light spectra of SNe Ia show a significant variation
of the composition and expansion velocities of the outermost layers,
whereas the post-maximum spectra are relatively uniform except for SN
1991bg.  The different \ni~ mass may produce the variation of the
spectra (\cite{Nugent-etal-95}).  Further analysis could provide the
abundance distribution of various elements (Fe, Ca, Si, Mg, O, C, and
others) in velocity space (\cite{Jeffery-etal-92}).

\noindent
{\em Light curves}: Recent high-quality charge-coupled device (CCD)
observations of SNe Ia have, on one hand, supported the basic
homogeneity of the optical light curve shape of SNe Ia. Good examples
include SNe 1980N and 1981D in the galaxy NGC 1316, whose light curve
shapes and brightnesses were almost identical.  On the other hand, the
observations have established significant variations of the maximum
brightness, the light curve shape, and their correlation (Fig.2).  In
theoretical models, the explosion energy goes into the kinetic energy
of expansion, $E$.  The light curves are powered by the radioactive
decay sequence $^{56}$Ni $\rightarrow$ $^{56}$Co $\rightarrow$
$^{56}$Fe.  The calculated light curve reaches its peak at about 15 to
20 days after the explosion and declines because of the increasing
transparency of the ejecta to gamma rays as well as the decreasing
input of radioactivity.  The light curve shape depends mainly on the
effective diffusion time $\tau_m \propto (\kappa M/v_{\rm
exp}c)^{1/2}$, where $\kappa$ is opacity, $c$ is the speed of light,
and \vv{exp}\ $\propto (E/M)^{1/2}$ (\cite{Arnett-96}).  For longer
$\tau_m$, the decline of the light curve is slower.  For larger
\mm{Ni}, the brightness is higher.  For the Ch models, typical values
of \mm{Ni} and $E$ are similar to those of W7, that is, \mm{Ni}\ = 0.6
$\pm$ 0.1 $M_\odot$ and $E$ = 1.3 $\pm$ 0.1 \e{44} J.

     The observed uniformity in the light curve implies a uniformity
of $M, E$, and \mm{Ni}.  A homogeneous $M$ and $E$ can be naturally
accounted for with the Ch models, although the central density can
vary with a small difference of mass near the Ch limit.  For the
sub-Ch models, the accretion process must somehow choose a relatively
narrow mass range, such as $M \approx$ 1.0 to 1.1 \ms.  The reported
variation of the maximum brightness of SNe Ia (\cite{Phillips-93},
\cite{Riess-etal-95}) may be a result of the variations of the
$^{56}$Ni mass.  The amount of \ni\ depends on (i) the flame speed in
the fast-deflagration models, (ii) the location and density of the
transition to a detonation in the delayed or late detonation models,
and (iii) the white dwarf mass in the sub-Ch models.  Because of
variations of these parameters, a variation of the \ni\ mass may be
possible.  For the Ch models, the dependence of optical opacities on
the temperature is important (\cite{Hoflich-etal-96}).  A smaller mass
of \ni\ means less heating of the surface layer, thus lowering
$\kappa$ and $\tau_m$.  This effect might explain the relation between
brightness and the decline rate of the light curve.  For the sub-Ch
models, \mm{Ni}\ varies approximately in proportion to $M$.  Then for
larger $M$, \mm{Ni}\ is larger and $\tau_m$ is longer; in other words,
brighter SNe Ia tend to decline more slowly (\cite{Arnett-96}).

\vskip 0.5cm

\centerline{\large\sf Nucleosynthesis in Galactic} 
\vskip 0.3cm

\centerline{\large\sf Chemical Evolution}

\vskip 0.5cm

\noindent
     Supernovae of different types have different progenitors, thus
producing different heavy elements on different time scales during the
chemical evolution of galaxies.  A reasonable mixture of the
heavy-element yields from SNe Ia and SNe II should be able to explain
the solar abundance pattern of heavy elements from O to the Fe group.
Nucleosynthesis products of SNe Ia and SNe II can be combined with
various ratios and compared with solar abundances of heavy elements
and their isotopes.  If the nucleosynthesis products of SNe II as a
function of stellar masses (\cite{Hashimoto-etal-89}) are adopted with
an upper mass limit of 50 \ms and the nucleosynthesis products of SNe
Ia are those from model W7, the best fit to known solar abundances is
obtained if the number of SNe Ia that have occurred relative to SNe II
is $N_{\rm Ia}/N_{\rm II}=0.12$ (\cite{Nomoto-etal-97}) (Fig. 4).
This is consistent with observation-based estimates that the SNe Ia
frequency is as low as 10 \% of total supernova occurrence
(\cite{vandenBergh-Tammann-91}).

     With this relative frequency, $^{56}$Fe from SNe Ia is about 50
\% of total $^{56}$Fe from all SNe.  For the Ch models, the abundance
ratios between neutron-rich species and $^{56}$Fe provide an important
constraints on the progenitor system.  The central density of the
white dwarf at thermonuclear runaway must be as low as \ltsim 2 \e9
\gmc to avoid overabundances of $^{58}$Ni, $^{54}$Cr, and $^{50}$Ti
relative to $^{56}$Fe, although the exact density depends on the flame
speed (\cite{Nomoto-etal-97}).  Such a low central density can be
realized by the accretion as fast as $\dot M >$ 1 $\times$ 10$^{-7}
M_\odot$ year$^{-1}$, which is consistent with the SD-Ch scenario.
SNe Ia yields thus contribute Fe enrichment to the chemical evolution
of the galaxies.  Because the progenitors of SNe Ia are low-mass stars
with relatively long lifetimes, Fe could be used as a clock to look
back into the timing of galaxy formation (\cite{Yoshii-etal-96}).

\vskip 0.5cm

\centerline{\large\sf Cosmological Parameters}

\vskip 0.5cm

\noindent
{\em Hubble constant}: SNe Ia are very good, but not perfect standard
candles.  The accuracy of the determination of \h0~ using SNe Ia
depends on (i) whether the peak absolute-magnitude dispersion of SNe
Ia is sufficiently small, and (ii) whether a precise
absolute-magnitude calibration of a SN Ia can be made.  Recent
progress has been made in the empirical determination of \h0.  For the
dispersion, the correlation between the maximum brightness and the
decline rate (or the light curve shape) is taken into account.
Multicolor light-curve shapes are also used for further corrections
(\cite{Riess-etal-95}).  This reduces the dispersion from 0.4 to 0.2
magnitudes in the Hubble diagram for SNe Ia (\cite{Riess-etal-95},
\cite{Hamuy-etal-95}) (Fig.5). For the calibration, the distances to
several host galaxies of SNe Ia (SNe 1895B, 1937C, 1960F, 1972E,
1981B, 1989B, and 1990N) have been determined from Hubble Space
Telescope observations of Cepheids in these galaxies
(\cite{Sandage-etal-95}).  On the basis of these data, $H_0$ is
estimated to be 58 $^{+7}_{-8}$ (\cite{Sandage-etal-95}), 63.1 $\pm$
6.3 (\cite{Hamuy-etal-95}), 64 $\pm$ 6 (\cite{Riess-etal-95}), and 55
$\pm$ 3 (\cite{Schaefer-95}) \kms Mpc$^{-1}$.

     Successful theoretical models can, in principle, give the
absolute maximum luminosities, thus providing good estimates of $H_0$
and \ome.  The maximum luminosity \ll{max}\ of SNe Ia when explained
with radioactive decay models has been used to estimate $H_0$
(\cite{Arnett-etal-85}).  From a comparison with individual SNe Ia,
$H_0$ = 67 $\pm$ 9 \kms Mpc$^{-1}$ (\cite{Hoflich-Khokhlov-96}), and
59 $\pm$ 13 \kms Mpc$^{-1}$ (\cite{Branch-etal-95}).  In addition, a
fit of the late-time synthetic spectra to observations gave $H_0$ = 68
$\pm$ 13 \kms Mpc$^{-1}$ (\cite{Ruiz-Lapuente-1996}).

These estimated values tend to converge to \h0~ $\approx$ 58 to 65
\kms Mpc$^{-1}$ where the error bars overlap.  The higher \h0~ is in
conflict with the age of the universe for \ome~$= 1$ and \omela $= 0$,
whereas the lower \h0~ could avoid this problem. Further efforts to
reduce the error bars will require the reexamination of sampling,
extinction corrections, and so on.

\noindent
{\em Density parameter}: In determining \ome~ and \omela~ from SNe Ia,
absolute calibration of the distances to SNe Ia is not needed, but
more accurate measurements of magnitudes and $z$ values are necessary.
In addition to the dispersion problem, whether a significant value of
\ome~ can be obtained depends on (i) whether a suitable sample of
remote SNe Ia can be obtained, and (ii) whether the peak luminosities
of SNe Ia are sufficiently free from the effects of cosmic and
galactic chemical evolutions.

     More than 60 SNe Ia with high $z$ values have been observed
(\cite{Perlmutter-etal-97}, \cite{Schmidt-etal-97}) (Fig. 5).  Values
derived from the first 7 supernovae at $z > 0.35$ are \ome~ =
$0.88^{+0.69}_{-0.60}$ for \omela~ = 0 or \ome~ $= 1 - $\omela = $0.94
^{+0.34}_{-0.28}$ for \ome~ + \omela~ = 1, assuming that evolutionary
effects are small.  To clarify whether the nature of high-$z$ SNe Ia
is the same as that of nearby SNe Ia, it is important to identify the
progenitors' evolution and population (\cite{Ruiz-etal-95}).
Systematic studies on the effects of metallicity of the progenitors
are underway.

\bigskip

\noindent
{\bf Figure 1:} Spectra of SNe Ia (SN 1994D and SN 1990N) about 1 week before 
maximum brightness.  SN1991T was peculiar (\cite{Filippenko-97}).
The y-axis is magnitude where the units of the flux $f_{\nu}$ are
ergs s$^{-1}$ cm$^{-2}$ Hz$^{-1}$. 

\noindent
{\bf Figure 2:} Empirical family of the visual light curves of SNe Ia,
which shows the brightness-decline rate relation.  The
triangles, squares, circles, and diamonds denote SNe 1991T, 1981B,
1986G, and 1991bg, respectively (\cite{Riess-etal-95}).

\noindent
{\bf Figure 3:} The non-LTE spectra of carbon deflagration model
W7 (unmixed) at 20 days and 23 days past explosion, which are 
compared with SNe 1992A and 1994D, respectively 
(\cite{Nugent-etal-97}). $F_{\lambda}$ is the flux in units of ergs 
s$^{-1}$ Angstrom$^{-1}$.

\noindent
{\bf Figure 4}: Solar abundance pattern based on synthesized heavy elements
from a composite of SNe Ia and SNe II with the most probable ratio
(\cite{Nomoto-etal-97}).  The dashed lines indicate typical
uncertainties of a factor of 2 involved in the observational and
theoretical abundances, where the abundance ratios are normalized to
unity for $^{56}$Fe.  The open and filled circles are used to clarify
the association to elements.

\noindent
{\bf Figure 5:} Hubble diagram for the seven high-$z$ SNe Ia out to $z
\sim$ 0.6, with some of the low-$z$ SNe Ia (\cite{Hamuy-etal-95}): (A)
uncorrected blue magnitudes $m_B$, and (B) $m_B$ with corrections of
the brightness-decline rate relation (\cite{Perlmutter-etal-97}).  The
solid curves are theoretical $m_B$ values for (\ome, \omela) $=$ (0,
0) (top curve), (1,0) (middle curve) , and (2,0) (bottom curve).  The
square points are not used in the analysis, because these points are
corrected based on the extrapolation outside the range of light curve
widths of low-redshift supernovae.  


\begin{thebibliography}{}
\bibitem[2]{Chandra} 
2. Ch is the maximum mass limit of the white
dwarfs in hydrostatic equilibrium, with a typical value of 1.46 \ms.  
For larger masses, pressures exerted by degenerate electrons
cannot sustain the gravity because of the special relativistic effect.
\end{thebibliography}
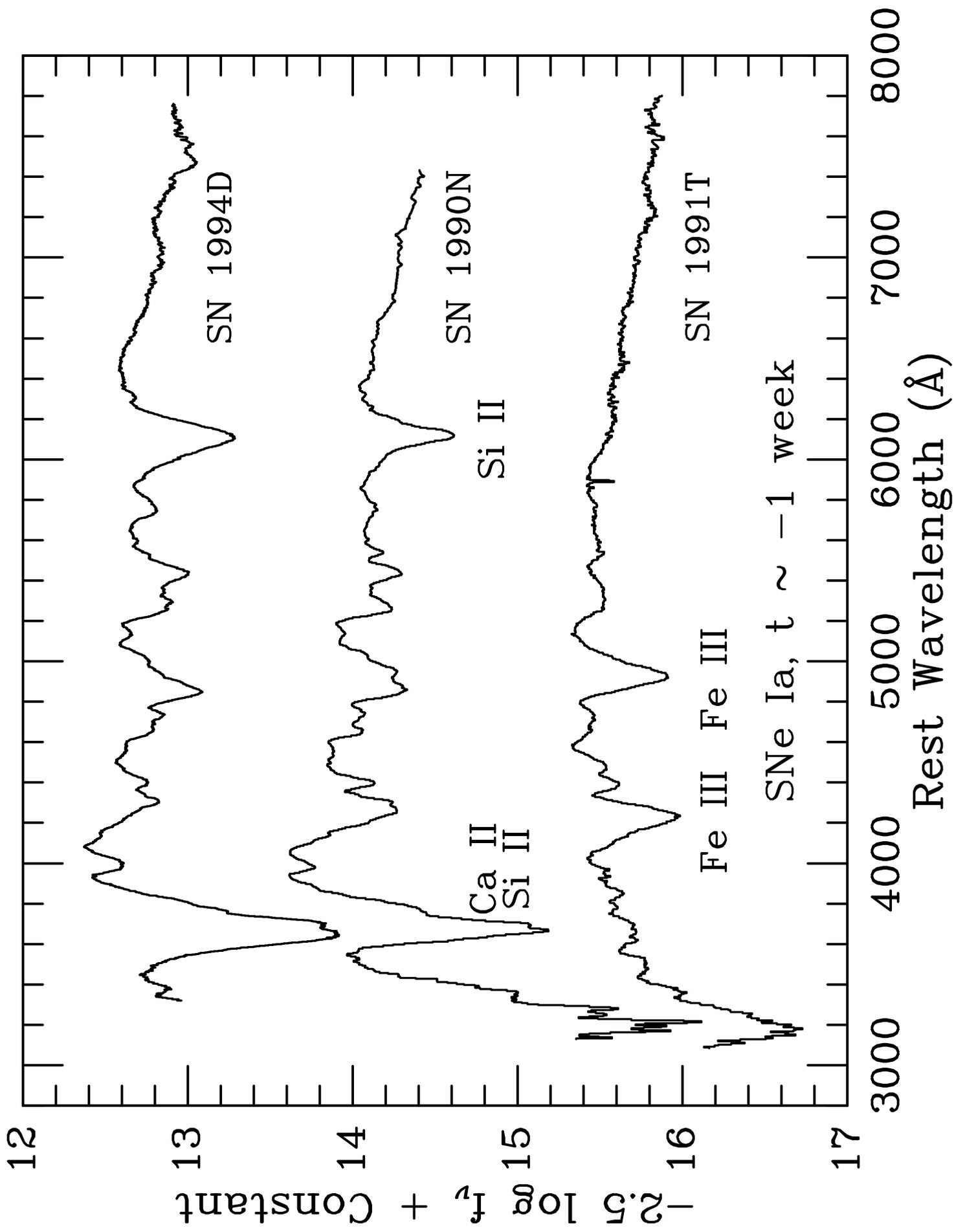
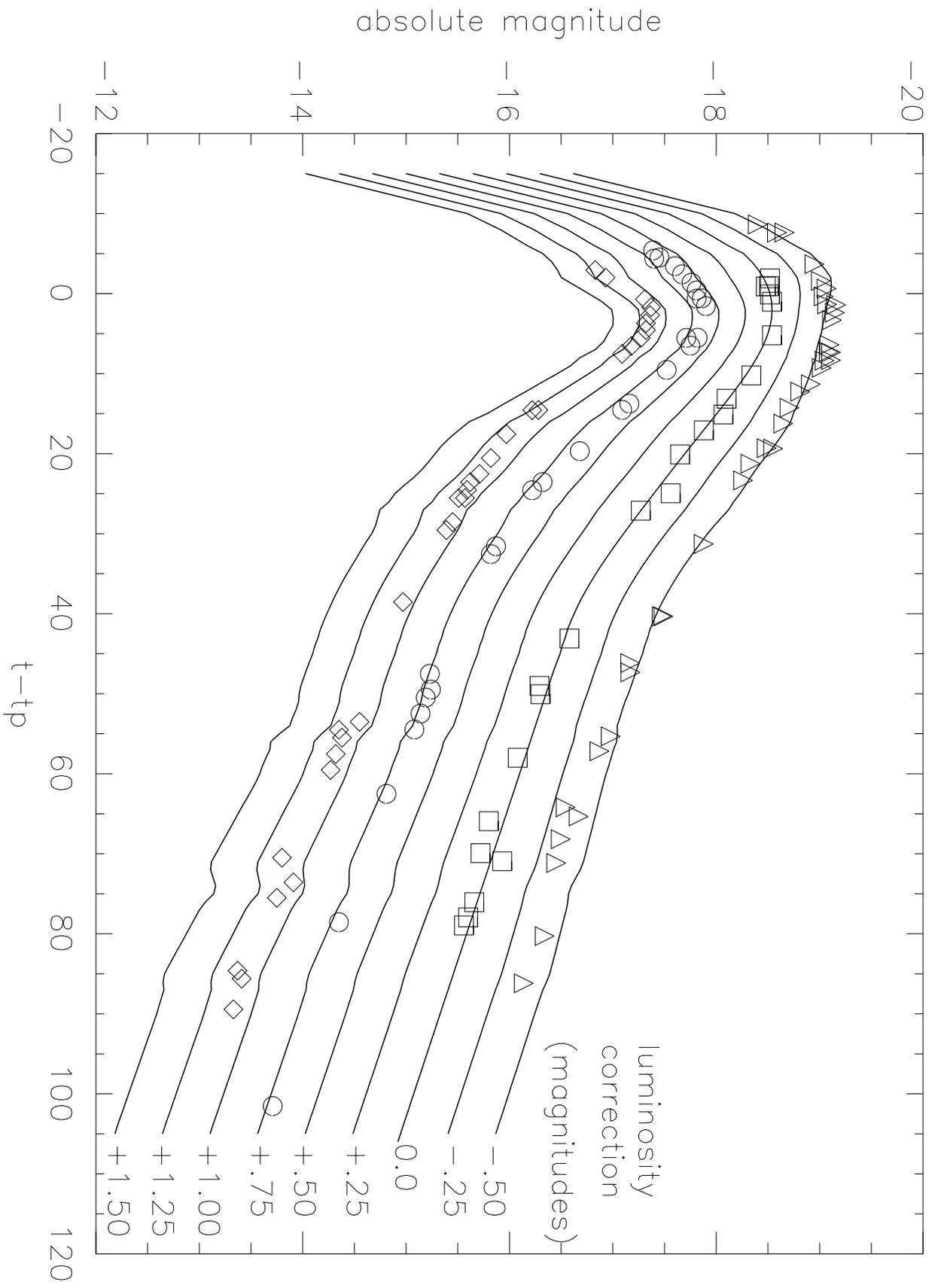
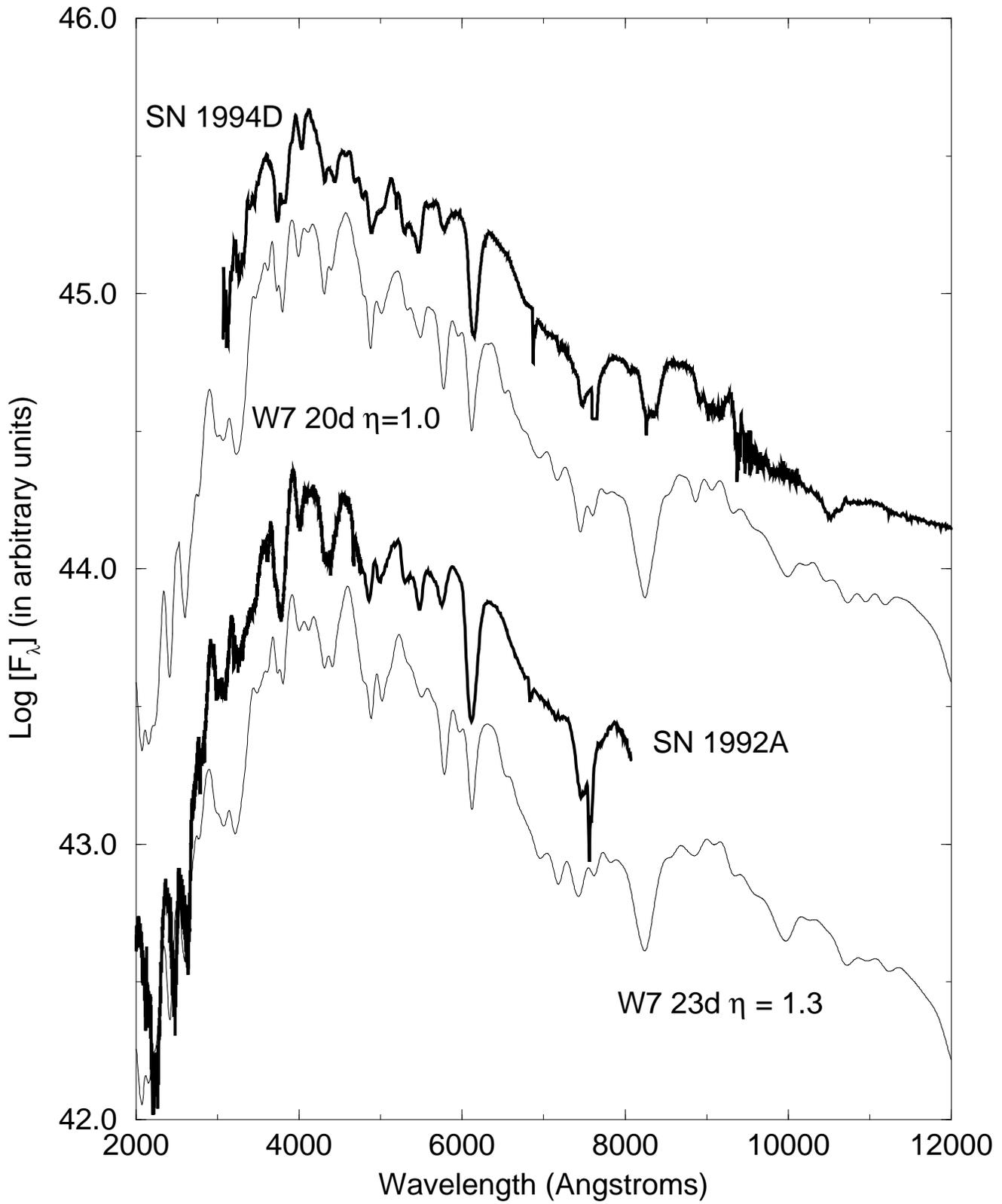
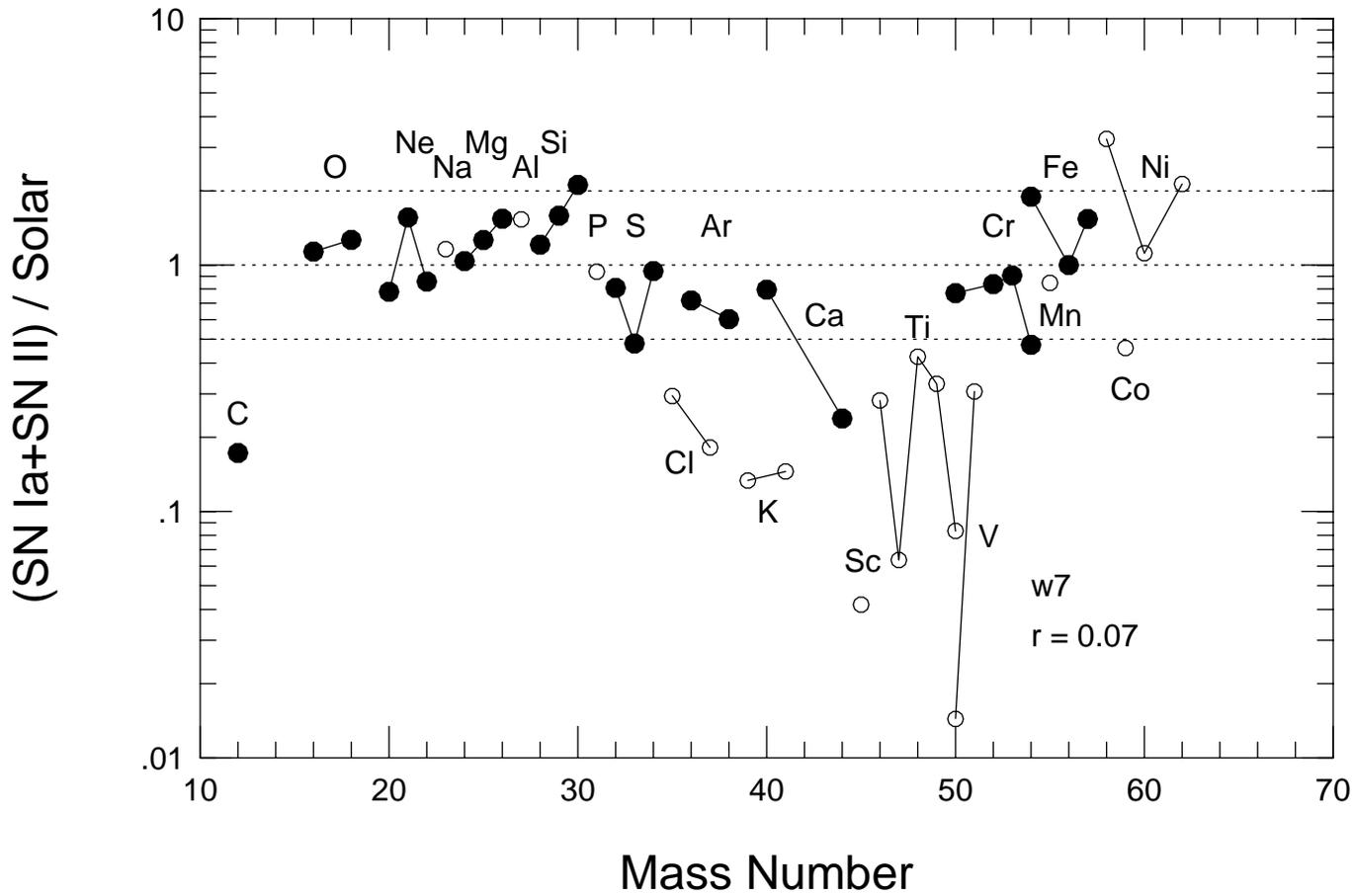
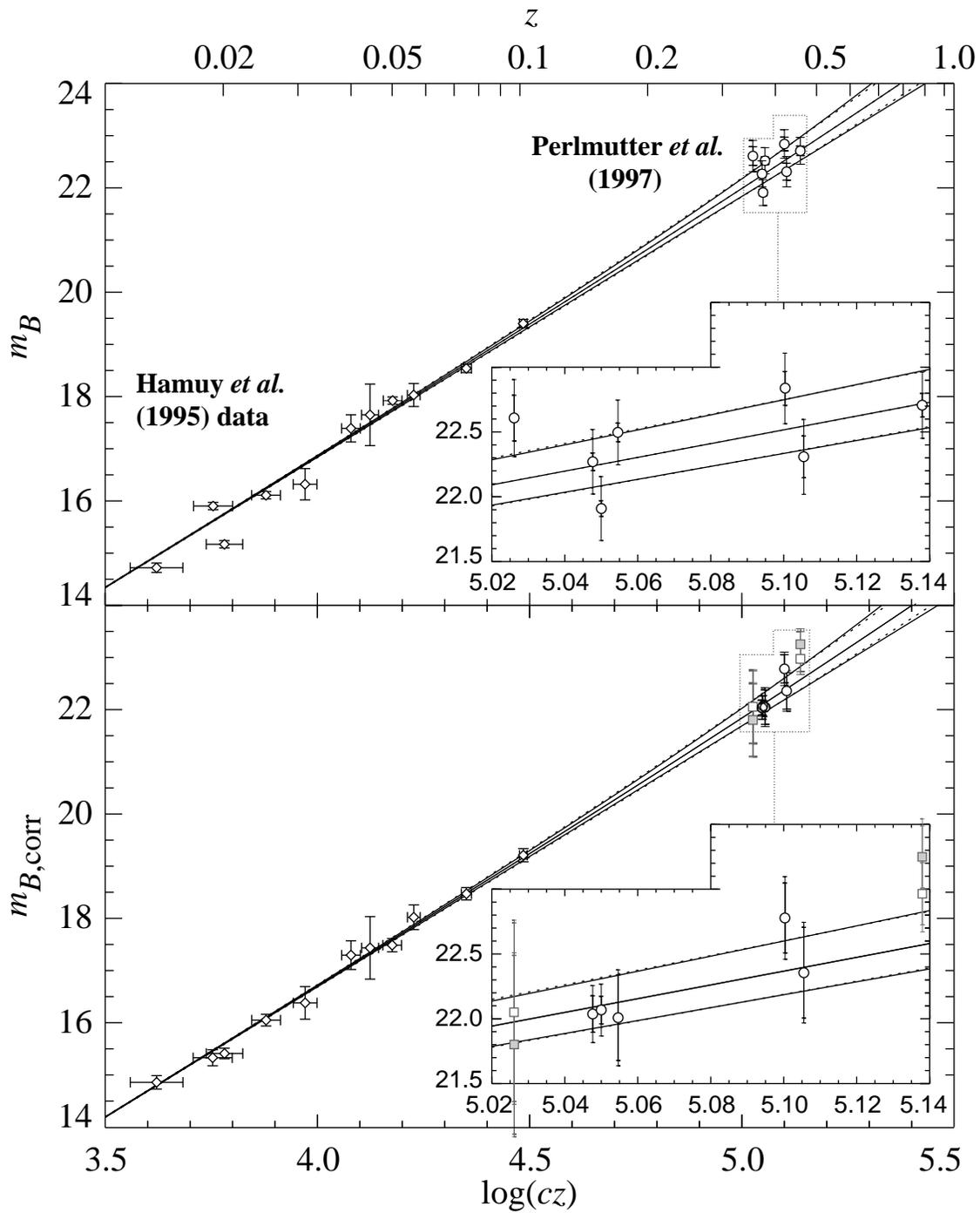
\end{document}